\documentclass[letterpaper,english,reprint,Appendixerscriptaddress,groupedaddress,pra,aps]{revtex4-1}
\usepackage[T1]{fontenc}
\usepackage[latin9]{inputenc}
\usepackage{color}
\usepackage{babel}
\usepackage{mathtools}
\usepackage{amsmath}
\usepackage{amssymb}
\usepackage{graphicx}
\usepackage{verbatim}
\usepackage[unicode=true,pdfusetitle,
 bookmarks=true,bookmarksnumbered=false,bookmarksopen=false,
 breaklinks=false,pdfborder={0 0 1},backref=false,colorlinks=false]
 {hyperref}
\hypersetup{
 colorlinks,linkcolor=red,anchorcolor=black,citecolor=blue}



\begin{document}
\preprint{APS/123-QED}
\title{Investigating the quench dynamics of the bound states in a spin-orbital
coupling system using a trapped ion}
\author{Hao-qing Zhang, Ming-zhong Ai, Jin-ming Cui}
\author{Yong-jian Han}
\email{smhan@ustc.edu.cn}
\author{Chuan-feng Li}
\author{Guang-can Guo}
\affiliation{CAS Key Laboratory of Quantum Information, University of Science and
Technology of China, Hefei, 230026, People Republic of China.and CAS
Center For Excellence in Quantum Information and Quantum Physics,University
of Science and Technology of China, Hefei, 230026, People Republic
of China.}

\date{\today}
\begin{abstract}
The quantum walk (QW), as the quantum analog of classical random walk, provides a feasible platform to study the topological phenomenon and non-equilibrium dynamics. Here, we propose a novel scheme to realize the quantum walk with a single trapped ion where the Fock states provides the walk space and zero phonon state $\left|n=0\right\rangle$ serves as its natural boundary. Thus, our scheme offers the unique opportunity to investigate the dynamics of the bound states of the corresponding topological systems. Particularly, the quench dynamics of the bound states can be extensively studied by tuning the bulk parameters and the local boundary operator, which are experimentally accessible. Our proposal not only offers a new approach to exploring the character of the bound states of the topological systems, but also offers a way to determine different phases through the dynamical processes.

\end{abstract}

\maketitle

\section{Introduction}

The topological matter has been studied extensively recently on different platforms \citep{von1986,hasan2010,qi2011,stormer1999,ando2013,wang2009,lu2014,lu2016,ozawa2019,aidelsburger2011,aidelsburger2013,cooper2019,goldman2016}. One of the interesting features is that the topological matter is protected by topology against local perturbations, such as the quantization of Hall conduct under impurity \citep{von1986,hasan2010,qi2011,stormer1999}.
The other unique characteristics of the topology matter are the appearance of the bound state at the boundary of the sample, for example, the open Su-Schrieffer-Heeger (SSH) chain in the one-dimensional system \citep{jackiw1976,su1979}, and robust edge state moving in one direction at the boundary of the two-dimensional system \citep{nakada1996,roulleau2008,atala2014,mancini2015,stuhl2015}.
The bulk topological invariants and the number of bound states can be connected by the bulk-edge correspondence \citep{hasan2010,qi2011,rudner2013}.

Though the equilibrium properties have been widely explored, the non-equilibrium dynamics of the topological system are still under investigation \citep{heyl2017,heyl2018,budich2016,yang2018,yang2018,wang2017,flaschner2018,tarnowski2019,floquet1883,shirley1965,grifoni1998,wilczek2012,choi2017,rovny2018,berges2004}. The quench process is the typical non-equilibrium process that has been studied in different topological systems \citep{xu2019,xu2020,flaschner2018,tarnowski2019}. On one hand, the bulk topological invariants, such as Chern number, defined on quantum states are known to be unchanged under unitary dynamics \citep{d2015,caio2015,hu2016,wang2017}, thus unchanged during the quench process. However, the non-unitary processes, such as the dissipation or decoherence process, will change the bulk topological invariants of the states during the quench process \citep{caio2015,hu2016,McGinley2018}.
Since the bulk-edge correspondence is only valid for the equilibrium situation \citep{hu2016,zhang2018,wang2019,yi2019},
the dynamics of the bound states at the boundary of the topological system is still elusive.
For example, Ref. \citep{caio2015} studies quench between topological and non-topological phases in Haldane model, while observing the presence or absence of edge modes. The dynamics of the bound states have attracted a lot of attention \citep{d2015}.In addition, how to experimentally observe the dynamics of the bound state is still an open question.

The quantum walk (QW) \cite{Aharonov2004}, which can be used to construct universal quantum
computation \citep{childs2009,childs2013,portugal2013}, has been
shown to be a powerful platform to study the equilibrium and non-equilibrium
topological properties of spin-orbital coupling systems \citep{Banuls2006,Kitagawa2010,Kitagawa2012,asboth2012,asboth2013,asboth2014,obuse2015,barkhofen2017,cardano2016,flurin2017,mochizuki2016,xiao2017,schreiber2010,crespi2013,schreiber2011}.
Particularly, it has been used to observe the bound states \citep{kitagawa2012observation,chen2018,nitsche2019}.
Different QWs have been experimentally realized in different platforms, such as photonics \citep{kitagawa2012observation,schreiber2010,broome2010,chen2018,xu2018,xu2019,xu2020},
neutral atoms \citep{karski2009,mugel2016}, superconductor
\citep{flurin2017,yan2019} and trapped ion \citep{xue2009,zahringer2010}. 

The trapped ions system, which can be accurately controlled and manipulated
\citep{leibfried2003,blatt2012}, is one of the most
ideal platforms for investigating quantum information processing and simulating non-equilibrium dynamics of many-body system \citep{eisert2015,zhang2017m,zhang2017t}.
Particularly, QW has been realized in one or two trapped $^{40}\text{Ca}^{+}$
ions in the phase space \citep{xue2009,zahringer2010}.
Here, we propose to encode QW onto the Fock states, which is similar to the result in Ref. \citep{oka2005}. The zero phonon state $\left|n=0\right\rangle $ acts as the natural boundary of the QW. With carefully designed laser sequences, the dynamics of the bound state can be experimentally investigated. We analyze the quench dynamics of the bound state by tuning different parameters
in this system.

The paper is organized as follows. In Sec. \ref{sec:2}, we briefly introduce the background of QW. Then, in Sec. \ref{sec:3}, we discuss how to realize the QW with a boundary in a trapped ion. In Sec. \ref{sec:4}, the corresponding between boundary operators and virtual bulk system are introduced. The main results are given in Sec. \ref{sec:5}: we first simulate the formation of the bound state with a novel scheme to verify the type of the bound state. Starting from the built bound state (if any), we study the dynamics of the bound state with quenched QW parameters and how quench rate effect edge population. Finally in Sec. \ref{sec:6}, we summarize the results.

\section{Quantum Walk Background\label{sec:2}}

\begin{figure*}
\mbox{%
\includegraphics[width=17.5cm]{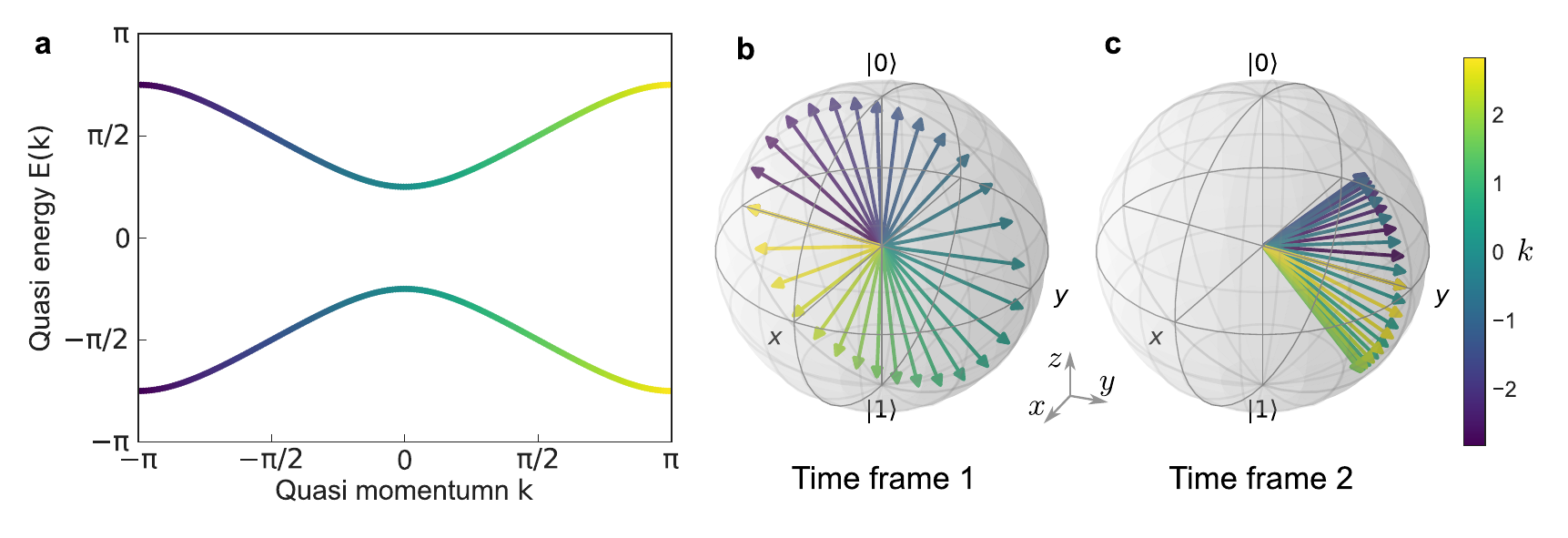}%
}\caption{\label{fig:split} Split-step quantum walk with the parameters $\theta_{1}=\frac{\pi}{2}$,
$\theta_{2}=0$. a. the dispersion relationship which is independent
of the time frame. b. and c. correspond to the two time frames
with CS to obtain the $\mathbb{Z}_{2}\times\mathbb{Z}_{2}$
topological invariants. In this case, $\nu^{\prime}=1$, $\nu^{\prime\prime}=0$
thus $\nu_{0}=1$, $\nu_{\pi}=0$ based on Eq. \eqref{eq:3}.}
\end{figure*}

QW is the quantum version of the classical random walk. The internal state (denoted by $\left|\uparrow\right\rangle ,\left|\downarrow\right\rangle $) and the position state (labeled as $\left|n\right\rangle $) of the walker in QW can be coupled through the
coin operation $R(\theta,n)$ ($\theta$ is the control parameter
and it may depend on the position of the walker $n$). We focus
on the split-step quantum walk \citep{asboth2012,asboth2013,asboth2014,Kitagawa2010,Kitagawa2012} (SSQW)  and its one-step Floquet operator
$U(\theta_{1},\theta_{2})$ is defined as: 
\begin{eqnarray}
U(\theta_{1},\theta_{2}) & = & S_{+}R(\theta_{2})S_{-}R(\theta_{1})\label{eq:1}
\end{eqnarray}
where $S_{\pm}$ is defined as $\sum_n\left|\text{\ensuremath{n\pm1}}\right\rangle \left\langle n\right|\otimes\left|\text{\ensuremath{\uparrow}}\right\rangle \left\langle \ensuremath{\uparrow}\right|+\mathbb{I}\otimes\left|\downarrow\right\rangle \left\langle \downarrow\right|$
and $R(\theta_k)=\sum_{n}\left|\text{\ensuremath{n}}\right\rangle \left\langle n\right|\otimes\mathrm{e}^{-\mathrm{i}\sigma_{y}\theta_k/2} (k=1,2)$ (the coin operator is independent of the position $n$). The quantum
state of the SSQW can be obtained by: $|\Psi\rangle_{n}=U^{n}(\theta_{1},\theta_{2})|\Psi\rangle_{0}$
where $|\Psi\rangle_{0}$ is the initial state of the evolution (generally,
we begin the evolution of the system with a localized product state). The effective
Hamiltonian of this periodical system can be derived from $U(\theta_{1},\theta_{2})=e^{-iH_{\mathrm{eff}}}$,
and $H_{\text{eff}}=\int_{k}E(k)\left|k\right\rangle \left\langle k\right|\otimes\bold{n}(k)\cdot \bold{\sigma}$
in momentum space. The dispersion relation and Bloch vector are calculated as below:

\begin{align}
\cos E(k) & =\cos\left(\frac{\theta_{2}}{2}\right)\cos\left(\frac{\theta_{1}}{2}\right)\cos k-\sin\left(\frac{\theta_{1}}{2}\right)\sin\left(\frac{\theta_{2}}{2}\right)\nonumber \\
n_{x}(k)= & \frac{\cos\left(\theta_{2}/2\right)\sin\left(\theta_{1}/2\right)\sin k}{\sin E(k)}\nonumber \\
n_{y}(k)= & \frac{\sin\left(\theta_{2}/2\right)\cos\left(\theta_{1}/2\right)+\cos\left(\theta_{2}/2\right)\sin\left(\theta_{1}/2\right)\cos k}{\sin E(k)}\nonumber \\
n_{z}(k)= & -\frac{\cos\left(\theta_{2}/2\right)\cos\left(\theta_{1}/2\right)\sin k}{\sin E(k)},
\end{align}
with $\bold{\sigma}=(\sigma_{x},\sigma_{y},\sigma_{z})$ are the Pauli matrices. The phase diagram of the QW can be determined by two different time frames with Chiral Symmetry \citep{asboth2013} (CS) as

\begin{align}
U_{1}(\theta_{1},\theta_{2})&=R\left(\theta_{1}/2\right)S_{+}R\left(\theta_{2}\right)S_{-}R\left(\theta_{1}/2\right),\nonumber
\\
U_{2}(\theta_{1},\theta_{2})&=R\left(\theta_{2}/2\right)S_{-}R(\theta_{1})S_{+}R\left(\theta_{2}/2\right).\label{eq:2}
\end{align}
And the different phases of the QW can be distinguished by a pair of $\mathbb{Z}_{2}\times\mathbb{Z}_{2}$
topological invariants \citep{asboth2013} (see Fig. \ref{fig:split}), where
\begin{eqnarray}
\nu_{0}&=\frac{1}{2}+\frac{1}{2}(\nu^{\prime}+\nu^{\prime\prime}),\nonumber \\
\nu_{\pi}&=\frac{1}{2}+\frac{1}{2}(\nu^{\prime}-\nu^{\prime\prime}),\label{eq:3}
\end{eqnarray}
where  $\nu_0, \nu_{\pi}$ are defined as the number of the bound state with eigenergy $0, \pi$ in the finite-size system, respectively, and $\nu^{\prime}$ ($\nu^{\prime\prime}$) is a bulk invariant which is defined as the winding number of system in the first (second)
time frame \cite{xu2020}(see Fig. \ref{fig:split}). Thus, Eq. \eqref{eq:3} clearly demonstrates the bulk-edge correspondence in one-dimensional QW system.

According to the bulk-edge correspondence, the bound states will appear at the boundary of the phases with different topology. Particularly, the vacuum can be viewed as a special phase and the bound state may appear at the boundary of a semi-finite (finite) QW system.

\section{EXPERIMENTAL PROPOSAL\label{sec:3}}
\begin{figure}
\includegraphics[width=8.5cm]{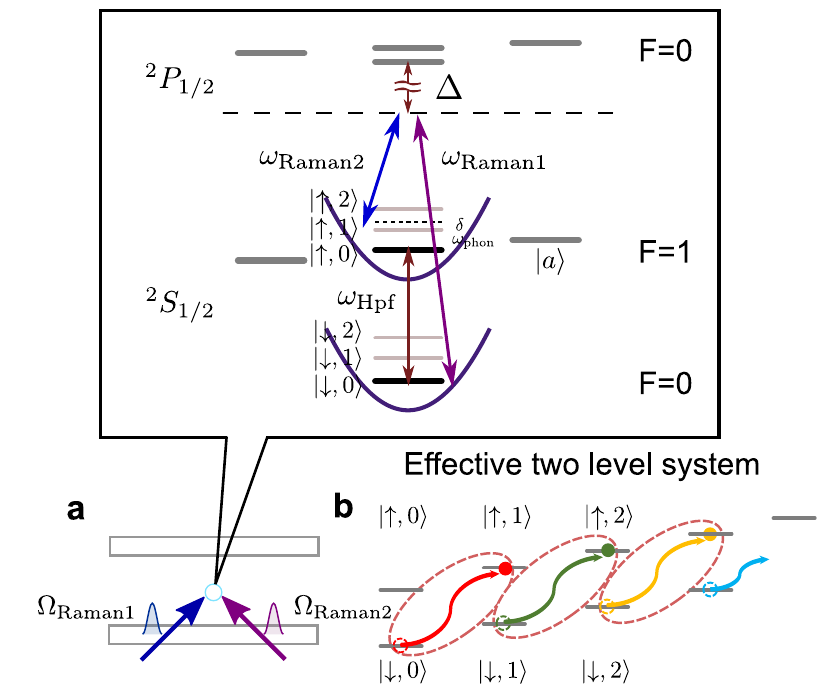}\caption{\label{fig:set}
The experimental sketches. a. Energy level diagram of single 
trapped $^{171}\text{Yb}^{+}$ ion. Perpendicular Raman beams are used to excite axial motional mode.  $\left|\uparrow,n\right\rangle$
and $\left|\downarrow,n\right\rangle$ represent for $\left|n\right\rangle \otimes\left|\uparrow\right\rangle$
and $\left|n\right\rangle \otimes\left|\downarrow\right\rangle$
with $\left|F=1,m_{F}=0\right\rangle \protect\coloneqq\left|\uparrow\right\rangle$
and $\left|F=0,m_{F}=0\right\rangle \protect\coloneqq\left|\downarrow\right\rangle$
of $^{2}S_{1/2}$ manifold, $|n\rangle$ for Fock state with
$n$ phonon. Auxiliary level $\left|F=1,m_{F}=1\right\rangle \coloneqq\left|a\right\rangle$ works for
temporal state shelving and it has no occupation after the whole operations. b. Effective two-level system is described by JC model, where $\left|\uparrow,n+1\right\rangle$ and $\left|\downarrow,n\right\rangle$ are coupled.}
\end{figure}
\begin{figure*}
\mbox{%
\includegraphics[width=17.5cm]{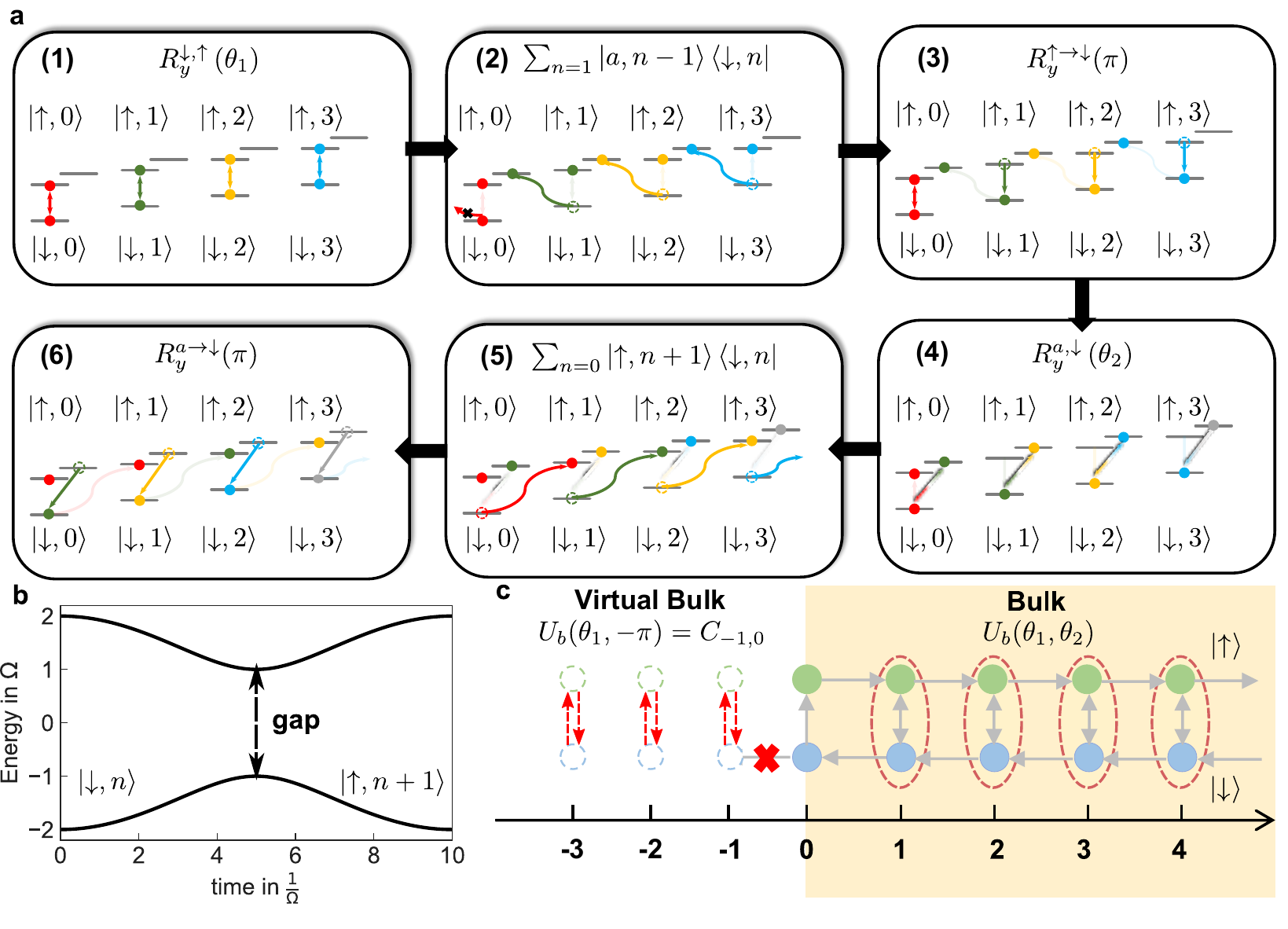}%
}\caption{\label{fig:experiment} a. Pulse sequences to realize QW with boundary. 
Six steps are required to construct $U(\theta_{1},\theta_{2})$ along the black arrow direction. Step one and four are coin operations which mix the spin states; the other steps are for spin-dependent phonon number shifting. $\left|\downarrow,0\right\rangle$ is blocked in step 2 for the phonon reduction, and then flips to $\left|\uparrow,0\right\rangle $ in step three. Thus it simulates the spin flipping operation at the boundary. Adiabatic processes in step two and five are used to fix the problem of inhomogeneous phonon shifting. b. Eigenergy of modulated Jaynes-Cummings model
$H(t)$ in subspace $\left|\downarrow,n\right\rangle \leftrightarrow\left|\uparrow,n+1\right\rangle $
(in the unit of $\Omega$). Rabi frequency $\Omega_0$ equals to detuning
$\delta_0$, while total transition time sets to be $\frac{10}{\Omega}$.
The adiabatic condition is satisfied when the evolution is much slower than the time scale set by the energy gap, then adiabatically transform the
state from $\left|\downarrow,n\right\rangle $ to $\left|\uparrow,n+1\right\rangle $
for all $n$. c. Virtual bulk topological invariant is defined with boundary
operation. With introducing virtual site at $n=-1$ lattice, we can
define the boundary "cut link" operator $C_{-1,0}$ (thus virtual bulk)}
\end{figure*}

Trapped ion systems have been proved as a powerful platform for quantum
simulation \cite{blatt2012}. Here, we propose a way to realize the QW with boundary by a single $^{171}\text{Yb}^{+}$ ion in a three-dimensional harmonic 
trap. In this system, the coin state of the QW is encoded in the $\left|F=1,m_{F}=0\right\rangle \coloneqq\left|\uparrow\right\rangle $
and $\left|F=0,m_{F}=0\right\rangle \coloneqq\left|\downarrow\right\rangle $
of $^{2}S_{1/2}$ hyperfine manifold of $^{171}\text{Yb}^{+}$ ion
with splitting $\omega_{\text{Hpf}}=2\pi\times12.6\ \text{\text{GHz}}$;
the lattice sites of the QW are encode in the number of the phonons,
where the zero phonon state $\left|n=0\right\rangle $ provides the natural boundary of the QW. See 
Fig. \ref{fig:set} for experimental sketchs of the proposal

To realize the QW, we need to implement two different basic operators
$R(\theta)$ and $S_{\pm}$. The rotation operator $R(\theta)$ of the coin in the QW is
easy to realize by manipulating the hyperfine state ($\left|\uparrow\right\rangle$
and $\left|\downarrow\right\rangle$) by microwave \citep{ospelkaus2011,leibfried2003} or by Stimulate-Raman-Process \citep{sorensen2006,leibfried2003}.
Implementing the operator $S_{\pm}$ in the Fock states needs more
work and is a bit complicated, where the auxiliary Zeeman energy level
$\left|F=1,m_{F}=1\right\rangle \coloneqq\left|a\right\rangle$ will be
introduced to provide a temporary shelving state. 

Generally, the interaction between the internal degree of freedom with energy splitting $\omega_{\text{Hpf}}$ and the phonons with frequency
$\omega_{\text{phon}}$ can be induced by a pair of proper selected
stimulated Raman beams with frequency $\omega_{\text{Raman1}}$ and
$\omega_{\text{Raman2}}$ (where $\omega_{\text{Raman1}}-\omega_{\text{Raman2}}=\omega_{\text{Hpf}}\pm\omega_{\text{phon}}+\delta$, $-$($+$) denote red (blue) sideband and $\delta$ is the two-photon detuning). $\Omega_{\text{Raman1}}$, $\Omega_{\text{Raman2}}$ donates for the Rabi frequencies which are proportional to the laser intensity of the Raman beams and $\Delta$ for single-photon detuning from the upper $^{2}P_{1/2}$ excited state. The induced interaction can be described by
the effective two-level Jaynes-Cummings Hamiltonian \citep{wu1997,lv2018}:
\begin{eqnarray}
H_{\text{JC}}=\frac{\Omega}{2}a\sigma_{+}e^{i\delta t}+\text{h.c.}\label{eq:jc}
\end{eqnarray}
under the rotation wave approximation (RWA), where $\Omega=\frac{\Omega_{\text{Raman1}}\Omega_{\text{Raman2}}}{2\Delta}$ is the
effective Rabi frequency, $a^{+}=\sum_{n=0}\sqrt{n+1}\left|n+1\right\rangle \left\langle n\right|$
is the creation operation of phonon, $\sigma_{+}(\sigma_{-})$
is the flipping operation $\left|\uparrow\right\rangle \left\langle \downarrow\right|$
($\left|\downarrow\right\rangle \left\langle \uparrow\right|$) of
spin.

With this integration, the number of phonons can be manipulated:
the red sideband beam gives the transition of $\left|n\right\rangle \otimes\left|\downarrow\right\rangle \leftrightarrow\left|n-1\right\rangle \otimes\left|\uparrow\right\rangle $
with Rabi frequency $\Omega_{n,n-1}=\sqrt{n}\Omega$ and the blue
sideband gives the transition of $|n\rangle\otimes\left|\downarrow\right\rangle\leftrightarrow|n+1\rangle\otimes\left|\uparrow\right\rangle$
with Rabi frequency $\Omega_{n,n+1}=\sqrt{n+1}\Omega$, where $\left|n\right\rangle $
is the Fock state with $n$ phonon. The controlled hopping between the state with different number of phonons is almost the same as the operator $S_{\pm}$, besides the Rabi frequency is dependent
on the number of the phonon (corresponding to the lattice site of the QW). The phonon
number dependence introduces additional complexity, as a result, to
implement the hopping operators $S_{\pm}$ homogeneously is the main
obstacle to realize the QW in this system. 

Actually, the homogeneous hopping operator $S_{\pm}$ indeed can be realized by
some adiabatic processes (called Stimulated Raman Adiabatic Passage
(STIRAP)) which has already been used to cool the motional state \citep{king1998,wunderlich2007,Gebert2016} and solve the inhomogeneous hopping problem \citep{um2016}. According to the adiabatic theorem \citep{simon1983}, when Hamiltonian of the system
changes slowly enough \citep{wunderlich2007}, the system will
keep on the $n$th eigenstate of the Hamiltonian $H(t)$ for all the time if
its initial state is the $n$th eigenstate of Hamiltonian $H(0)$ and the $n$th eigenstate is isolated from the others. 

Based on the adiabatic theorem mentioned above, we can set $\Omega(t)=\Omega_0\sin(\frac{\pi t}{\tau})$
and $\delta(t)=\delta_0\cos(\frac{\pi t}{\tau})$ ($\tau$ is the total
operation time) in Eq. \eqref{eq:jc} to construct
the time dependent Hamiltonian $H(t)$ in the adiabatic process. The
initial state $\left|n\right\rangle \otimes\left|\downarrow\right\rangle$
is an eigenstate of the Hamiltonian $H(0)$. When
the parameter satisfies the adiabatic condition, i.e. $\frac{d\theta(t)}{dt}\ll\sqrt{\Omega(t)^2+\delta(t)^2}$
($\tan\theta(t)=\frac{\Omega(t)}{\delta(t)}$ is the angle on the
Bloch sphere during the evolution), the system will stay at the corresponding eigenstate
of $H(t)$. The adiabatic process is not dependent on the number
of the phonon if the order of the eigenstates is given, see Fig.
\ref{fig:experiment}b) for eigenergy with given parameters during the whole adiabatic process. In order to speed up the adiabatic process, further methods to suppress the non-adiabatic excitation \citep{um2016} and to reshape the waveform for shortcut Raman passage have been proposed and realized in the experiment \citep{Bergmann1998,Berry2009,chen2010,du2016}. 

Therefore, based on the STIRAP method, the key operators $R(\theta)$
and $S_{\pm}$ for the QW can be realized with a trapped ion and the
whole QW evolution $U(\theta_{1},\theta_{2})$ can be realized by
the following six steps (for convenience, we define $\Delta\omega=\omega_{\text{Raman1}}-\omega_{\text{Raman2}}$, Zeeman splitting under magnetic field $\omega_{\text{Zm}}$ and $R_y(\theta)=e^{-i\sigma_y\theta/2}$):

\begin{enumerate}
\item Apply rotation $R_{y}(\theta_{1})$ in the spin state space (spanned $\left|\uparrow\right\rangle $
and $\left|\downarrow\right\rangle $), which can be easily realized
by two Raman laser with $\Delta\omega=\omega_{\text{Hpf}}$. The corresponding
evolution $R_{y}(\theta_{1})=\sum_{n=0}^{\infty}\left|n\right\rangle \left\langle n\right|\otimes[\cos(\theta_{1}/2)(\left|\uparrow\right\rangle \left\langle \uparrow\right|+\left|\downarrow\right\rangle \left\langle \downarrow\right|)-\sin(\theta_{1}/2)(\left|\uparrow\right\rangle \left\langle \downarrow\right|-\left|\downarrow\right\rangle \left\langle \uparrow\right|)+\left|a\right\rangle \left\langle a\right|]$
is independent of the number of the phonon and rotation angle $\theta_{1}$ can be controlled precisely by pulse duration;
 
\item Apply STIRAP for first red sideband (spanned $\left|a\right\rangle $ and $\left|\downarrow\right\rangle$), the frequency of the two Raman
laser are chosen as $\Delta\omega=\omega_{\text{Hpf}}+\omega_{\text{Zm}}-\omega_{\text{Phon}}+\delta(t)$
and the evolution can be effectively written as $S_{-}=\left|0\right\rangle \left\langle 0\right|\otimes\left|\downarrow\right\rangle \left\langle \downarrow\right|-\sum_{n=1}^{\infty}\left|n-1\right\rangle \left\langle n\right|\otimes\left|a\right\rangle \left\langle \downarrow\right|+\sum_{n=0}^{\infty}\left|n\right\rangle \left\langle n\right|\otimes\left|\uparrow\right\rangle \left\langle \uparrow\right|$.
Notice $\left|0\right\rangle\otimes\left|\downarrow\right\rangle$ can not be driven at this step;

\item Apply $R_{y}(\pi)$ (spanned $\left|\uparrow\right\rangle $
and $\left|\downarrow\right\rangle $) as in step 1, the evolution
is: $\sum_{n=0}^{\infty}\left|n\right\rangle \left\langle n\right|\otimes(\left|\downarrow\right\rangle \left\langle \uparrow\right|-\left|\uparrow\right\rangle \left\langle \downarrow\right|+\left|a\right\rangle \left\langle a\right|)$;

\item Apply $R_{y}(\theta_2)$ in (spanned $\left|a\right\rangle $
and $\left|\downarrow\right\rangle $) as in the step 1 with $\Delta\omega=\omega_{\text{Hpf}}+\omega_{\text{Zm}}$,
the evolution is: $\sum_{n=0}^{\infty}\left|n\right\rangle \left\langle n\right|\otimes[\cos(\theta_{2}/2)(\left|\downarrow\right\rangle \left\langle \downarrow\right|+\left|a\right\rangle \left\langle a\right|)-\sin(\theta_{2}/2)(\left|\downarrow\right\rangle \left\langle a\right|-\left|a\right\rangle \left\langle \downarrow\right|)+\left|\uparrow\right\rangle \left\langle \uparrow\right|]$;

\item Apply STIRAP for the first red sideband (spanned $\left|\downarrow\right\rangle $
and $\left|\uparrow\right\rangle $) with $\Delta\omega=\omega_{\text{Hpf}}+\omega_{\text{Zm}}+\omega_{\text{phon}}+\delta(t)$,
the effective evolution is: $S_{+}=-\sum_{n=0}^{\infty}\left|n+1\right\rangle \left\langle n\right|\otimes\left|\uparrow\right\rangle \left\langle \downarrow\right|+\left|n\right\rangle \left\langle n\right|\otimes\left|a\right\rangle \left\langle a\right|$;

\item Apply $R_{y}(\pi)$ (spanned  $\left|a\right\rangle $
and $\left|\downarrow\right\rangle$) with $\Delta\omega=\omega_{\text{Hpf}}+\omega_{\text{Zm}}$,
the evolution is: $\sum_{n=0}^{\infty}\left|n\right\rangle \left\langle n\right|\otimes(-\left|\downarrow\right\rangle \left\langle a\right|+\left|a\right\rangle \left\langle \downarrow\right|+\left|\uparrow\right\rangle \left\langle \uparrow\right|)$;
\end{enumerate}
the whole process is clearly shown in Fig. \ref{fig:experiment}a). To complete
the QW and observe the physical phenomenon, we need to repeat the whole cycle many times. It is emphasized
that the auxiliary state $\left|a\right\rangle $ which is introduced
for temporal shelving and is empty when the six-step cycle is completed.
It means there is no probability to detect the ion at the state $\left|a\right\rangle$
and no information is leaked. 
Obviously, the zero phonon state $\left|n=0\right\rangle $ which provides the boundary in the
QW is special (see above step 2), we can rewrite the evolution operators
$U(\theta_{1},\theta_{2})$ to distinguish the boundary site from
the others as below (note a minus sign appear because of $\left|\downarrow\right\rangle \rightarrow\left|\uparrow\right\rangle \rightarrow\left|\downarrow\right\rangle$ gain an additional sign):
\begin{align}
U_{x>0}(\theta_{1},\theta_{2}) & =S_{x>0}^{+}e^{-i\theta_{2}\sigma_{y}/2}S_{x>0}^{-}e^{-i\theta_{1}\sigma_{y}/2},
\end{align}
where:
\begin{align}
S_{x>0}^{+} & =\sum_{n=0}^{\infty}\left|n\right\rangle \left\langle n\right|\otimes\left|\downarrow\right\rangle \left\langle \downarrow\right|-\left|n+1\right\rangle \left\langle n\right|\otimes\left|\uparrow\right\rangle \left\langle \uparrow\right|\label{eq:5}\\ \nonumber
S_{x>0}^{-} & =\sum_{n=0}^{\infty}\left|n\right\rangle \left\langle n\right|\otimes\left|\uparrow\right\rangle \left\langle \uparrow\right|-\left|n\right\rangle \left\langle n+1\right|\otimes\left|\downarrow\right\rangle \left\langle \downarrow\right|; 
\end{align}
and the boundary operator:
\begin{equation}
U_{x=0}(\theta_{1})=e^{i\phi}\left|0\right\rangle \left\langle 0\right|\otimes\left|\uparrow\right\rangle \left\langle \downarrow\right|e^{-i\theta_{1}\sigma_{y}/2},\label{eq:6}
\end{equation}
the parameter $\phi$ could be controlled after step 2 in our experimental proposal: only $\left|0\right\rangle \otimes\left|\downarrow\right\rangle$ state has non-zero occupation among all $\left|\downarrow\right\rangle$ states. Thus $\sigma_z$ operation between $\left|\downarrow\right\rangle$ and other energy levels (except $\left|\uparrow\right\rangle$ and $\left|a\right\rangle$) gives a relative phase for $\left|n\right\rangle \otimes\left|\uparrow\right\rangle$ compared with other sites. Due to the Chiral Symmetry and the Particle-Hole Symmetry (PHS) requirement of the QW, $\phi$ can only be $0$ or $\pi$.

We further discuss how to realize two-dimensional QW \citep{Kitagawa2010,chen2018,yan2019,sajid2019,li2020},
which could be associated with non-trivial Chern numbers and the more
complex Floquet band structure. Single-step with two coin operations and two walk operations is shown below:
\begin{eqnarray}
U=S_{y}R(\theta_{2})S_{x}R(\theta_{1}),
\end{eqnarray}
where $S_{x}=\sum_{x,y}\left|x+1,y\right\rangle \left\langle x,y\right|\otimes\left|\uparrow\right\rangle \left\langle \uparrow\right|+\left|x-1,y\right\rangle \left\langle x,y\right|\otimes\left|\downarrow\right\rangle \left\langle \downarrow\right|$
and $S_{y}=\sum_{x,y}\left|x,y+1\right\rangle \left\langle x,y\right|\otimes\left|\uparrow\right\rangle \left\langle \uparrow\right|+\left|x,y-1\right\rangle \left\langle x,y\right|\otimes\left|\downarrow\right\rangle \left\langle \downarrow\right|$,
$x,y$ are independent freedoms. In our proposal,
two dimensions of the particle propagation can be encoded in two different motional
modes of the ion: for example, the axis and radial motional
modes with the resolved frequency $\omega_{z},\omega_{r}$.
Additional Raman beams are required to excite the motional modes.
Of course, this 2D model has two reflecting boundaries for $x,y\geq0$. Higher-dimensional QW is also possible with more resolved motional modes involved. 

With this setup, we can investigate the dynamics of the bound state in the topological system.

\section{Boundary and Virtual Bulk phase \label{sec:4}}
Based on the bulk-edge correspondence theory, some bound states will
appear at the interface of two topologically different bulk phases.
Particularly, the bound state may appear at the boundary of a finite
or semi-infinite topological system, which can be viewed as the interface
of the real bulk topological phase and some virtual bulk phase. Interestingly,
the virtual bulk phase is only dependent on the local operator of the boundary. To clearly establish the relations among the bound states, the bulk topological invariants, and the boundary condition, we need to define the correspondence between the virtual bulk phases and the local boundary operators. 

To establish the correspondence, we map our semi-infinite model to the "cut link" model suggested in \citep{asboth2013}.
In the "cut link" model, the shift operators of the "uncut link" ($S_{n,n+1}$)
and "cut link" operation ($C_{n,n+1}$) between the site $n$ and $n+1$ is introduced
as:
\begin{align}
S_{n,n+1}= & \left|n\right\rangle \left\langle n+1\right|\otimes\left|\downarrow\right\rangle \left\langle \downarrow\right|+\left|n+1\right\rangle \left\langle n\right|\otimes\left|\uparrow\right\rangle \left\langle \uparrow\right|\nonumber \\
C_{n,n+1}= & \left|n+1\right\rangle \left\langle n+1\right|\otimes\left|\uparrow\right\rangle \left\langle \downarrow\right|-\left|n\right\rangle \left\langle n\right|\otimes\left|\downarrow\right\rangle \left\langle \uparrow\right|,
\end{align}
the standard SSQW in Eq. \eqref{eq:1} ($-\infty \le n \le \infty$) could be decomposed with these new operators as
\begin{equation}
    U(\theta_{1},\theta_{2})=\sum(\cos(\frac{\theta_{2}}{2})S_{n,n+1}+\sin(\frac{\theta_{2}}{2})C_{n,n+1})R(\theta_{1})  \label{eq:new}
\end{equation}

Particularly, in the semi-infinite system, the lattice space stops at $n=0$ site, however, for convenience, we can still introduce an additional virtual site $n=-1$ as shown in Fig. \ref{fig:experiment}c). The "cut link" operator $C_{0,-1}$ which serves as the boundary condition and does not affect the evolution of our system. In the following we can see that the operator $C_{0,-1}$, which can be experimentally controlled, plays the key role in the emergence of the bound state.

Similar to Eq. \eqref{eq:new} in the standard SSQW model, we can rewrite Eq. \eqref{eq:5} and \eqref{eq:6} in the semi-infinite model of our proposal with "uncut link" and "cut link" operators. Take $\phi=0$ as an example:
\begin{align}
U_{b(x>0)}(\theta_{1},\theta_{2}) & =-\sum_{n=0}(\cos(\frac{\theta_{2}}{2})S_{n,n+1}+\sin(\frac{\theta_{2}}{2})C_{n,n+1})R(\theta_{1})\nonumber \\
U_{x=0}(\theta_{1}) & =C_{0,-1}R(\theta_{1}),
\end{align}
obviously, $U_{x=0}(\theta_{1})$ is directly connected by the operation $C_{0,-1}$.

Comparing the bulk operator $U_b(\theta_1,\theta_{2})$ and the boundary operator $U_{x=0}(\theta_{1})$, we can find that
$U_b(\theta_1,\theta_2=-\pi)=U_{x=0}(\theta_{1})$. As a result, the boundary ($n=0$) can be viewed as the interface of two bulks: one is the real bulk system with bulk operator $U_b(\theta_1,\theta_{2})$, and the other is the virtual bulk system with the bulk operator $U_b(\theta_1,-\pi)$.

\begin{figure}
\includegraphics[width=8.5cm]{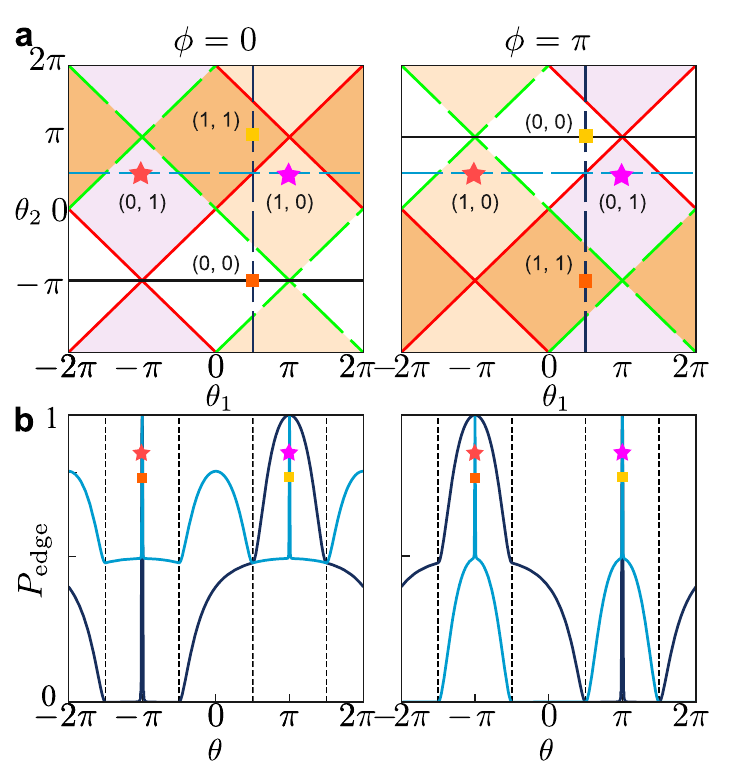}\caption{\label{fig:phase diagram} The phase diagram of SSQW with boundary and simulated
localization probability at the boundary under different parameters. a. Phase diagram
of $\phi=0,\pi$ (left and right diagrams). The virtual bulk phase $(0,0)$ is
defined by "cut link" operation $C_{-1,0}$, which
corresponds to the region: $\theta_{2}=-\pi$ ($\pi$) for $\phi=0$ ($\pi$)
as shown by the black solid line. In the phase diagram, the dashed green lines mean gap
close at $E=0$ and the solid red lines mean gap close at $E=\pi$. The whole diagram is divided into four different
phases. b. simulation results after 100 steps with $\phi=0,\pi$ (left and right diagrams).
The initial state is prepared as $\left|0\right\rangle \otimes\left|\downarrow\right\rangle $.
The appearance of bound states is represented by $P_{\text{edge}}$. The
cyan and blue curves correspond to scan over fixed $\theta_{1}=\pi/2$ and
$\theta_{2}=\pi/2$. Sharp boundary between different regions could be observed for the number of $E=0,\pi$ bound states changing.}
\end{figure}

With the corresponding of the boundary operator and the virtual bulk operator, we can obtain the phase diagram of the virtual bulk system. The parameter $\theta_2$ has only two values, $-\pi$ (corresponding to $\phi=0$ in boundary operator $U_{x=0}$) and $\pi$ (corresponding to $\phi=\pi$ in boundary operator $U_{x=0}$), due to the CS requirement. Consequently, the phase diagram of the virtual bulk system include two lines in Fig. \ref{fig:phase diagram}a).

If and only if the virtual bulk system and the real bulk system have different topology, the bound states can appear. To clearly verify the statement, we prepare the QW system onto the state $\left|0\right\rangle \otimes\left|\downarrow\right\rangle$ and take $\phi=0$ as an example. The semi-finite system are subsequently evolved under unitary operator $U(\theta_{1},\theta_{2})$. We first consider the special case with $\theta_{1}=\pi/2,\theta_{2}=-\pi$ (red star in phase $(0,1)$ in Fig. \ref{fig:phase diagram}a), where the system is evolved as: 
\begin{align}
\left|0\right\rangle \otimes\left|\downarrow\right\rangle  & \stackrel{1}{\rightarrow}-\frac{1}{\sqrt{2}}(\left|0\right\rangle \otimes\left|\uparrow\right\rangle +\left|1\right\rangle \otimes\left|\downarrow\right\rangle) \stackrel{2}{\rightarrow}-\left|0\right\rangle \otimes\left|\downarrow\right\rangle .
\end{align}
It is clear, after two steps the particle returns to its initial state with an
additional minus sign which indicate that bound state is the eigenstate with eigenvalue $E=\pi/2$ \citep{Kitagawa2012,asboth2012} ( the pink star in the phase $(1,0)$ and orange square in the phase $(0,0)$). 
While for the parameter $\theta_{1}=\pi/2,\theta_{2}=\pi$
(yellow square in the phase $(1,1)$), similar analyses conclude that the other bound state (eigenstate) with $E=0,\pi$ exist, i.e.
\begin{equation}
\left|0\right\rangle \otimes\left|\downarrow\right\rangle \stackrel{1}{\rightarrow}\left|0\right\rangle \otimes\frac{1}{\sqrt{2}}(\left|\uparrow\right\rangle +\left|\downarrow\right\rangle )\stackrel{2}{\rightarrow}\left|0\right\rangle \otimes\left|\downarrow\right\rangle .
\end{equation}

For more general cases, we use the localization probability at the boundary: $P_{\text{edge}}=p_0+p_1$, where $p_0$, $p_1$ are the probability at $n=0, 1$ site respectively \cite{barkhofen2017}, as the indicator of the emergence of the bound state. When bound states appear, the probability $P_{\text{edge}}$ will be stable with a nonzero value along the evolution steps, and the stable on-site probability $P_n= _N\langle \Psi|n|\Psi \rangle_N$ ($N\rightarrow \infty$) will exponentially decay along with $n$ for the left finite sites, which means the $P_n\propto e^{-n/\lambda}$ and $\lambda$ is the localization length. 

The localization probability $P_{\text{edge}}$  after 100 steps verse different parameters is shown in Fig. \ref{fig:phase diagram}b). In the left of Fig. \ref{fig:phase diagram}b), the parameter $\theta_2$ of the virtual bulk system is chosen as $\theta_2=-\pi$ (set the control parameter $\phi=0$ in our experimental proposal), the parameter $\theta_2$ in the real bulk system is fixed to $\pi/2$, and the parameter $\theta_1$ is the same in the virtual and the real bulk system. The results are shown as cyan curve of Fig. \ref{fig:phase diagram}b). We scan the parameter $\theta_1$ in one period. We study in the time frame $R(\theta_{1}/2)S_{+}R(\theta_{2})S_{-}R(\theta_{1}/2)$ which preserves CS. In this case, the topology of the real bulk system and the virtual bulk system are always different and, subsequently, the bound state always exists. We also fixed $\theta_1=\pi/2$ (in the real and the virtual bulk system) and scan $\theta_2$ in the real bulk system ($\theta_2=-\pi$ in the virtual bulk system), the results are shown as the blue curve of Fig. \ref{fig:phase diagram}b), from which we can see that there are no bound states at the region $\theta_2 \in [-3\pi/2,-\pi/2]$ for the topology of the real and the virtual bulk system are the same. The similar results when we fix the parameter $\theta_2 $ in the virtual bulk system to $\pi$ are shown in the right of Fig. \ref{fig:phase diagram} (set the control parameter $\phi=\pi$ in our experimental proposal). 

All the calculations are consistent with our observation: the bound states appear if and only if the topology of the virtual bulk is different from the real bulk system.

\section{DYNAMICS OF BOUND STATE\label{sec:5}}

\begin{figure}
\includegraphics[width=8.5cm]{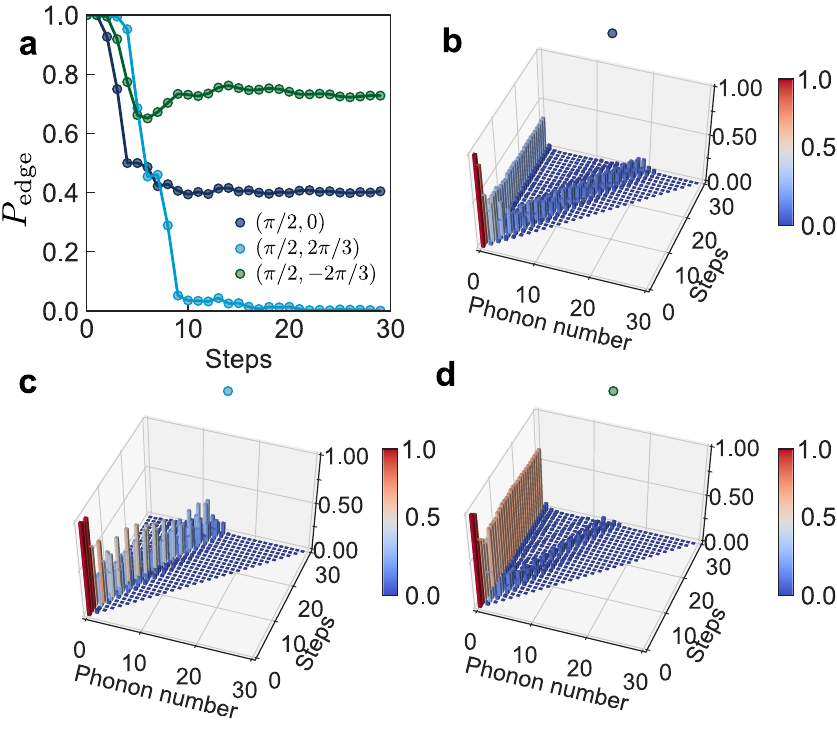}\caption{\label{fig:bound state}Existence of bound state in different regions. We study
points in different topology: blue: $(\pi/2,0)$, cyan: $(\pi/2,-2\pi/3)$, green: $(\pi/2,2\pi/3)$ for the rotation angle in the real bulk system while the fixed the virtual bulk system as $\theta_2=-\pi$. The particle is prepared as $\left|0\right\rangle \otimes\left|\downarrow\right\rangle$.
a. Edge population $P_{\text{edge}}$ verse steps, we can
observe that $P_{\text{edge}}$ decays to near zero after about 10 steps
when the real bulk system in the trivial phase. We further monitor $P_{\text{edge}}$ verse steps and phonon numbers with real bulk system in b. $(0,1)$, c. $(0,0)$, d. $(1,1)$, which select the same parameters as in a.
The clear bound states are found at the boundary when the real bulk systems are in the non-trivial phase as expected.}
\end{figure}

\begin{figure}
\includegraphics[width=8.5cm]{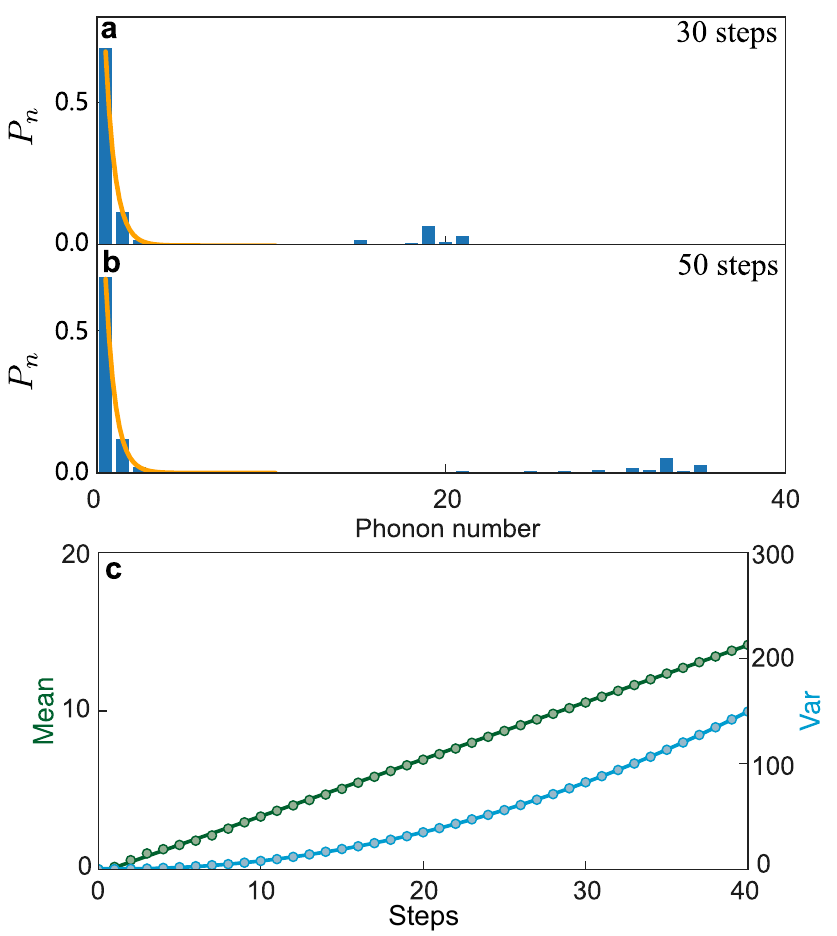}\caption{\label{fig:phonon distribution}The simulation result of the phonon state distribution after a certain steps, with $\theta_{1}=\pi/2,\theta_{2}=0$ and $\phi=0$. A single $E=0$ bound state appears as prediction for real bulk system in the phase $(1,0)$ has the different topology from the virtual bulk system. a. and b. the result after 30 and 50 steps respectively. The histogram shows the phonon distribution and the orange line is the analytical solution of the eigenstate. c. the mean and variance of QW verse steps. 
The feature of quadratic increment of the deviation and linear increment of the mean are the patterns of acceleration of the classical random walk.}
\end{figure}

In the previous section, we have verified the existence of bound
states when the topology of the virtual bulk system and the real bulk system are different. In this section, we will study the dynamics of the bound states. The dynamics of the bound states include two different situations: one is the dynamics of the formation of the bound states from a local initial state during the evolution; the other is the dynamics of the bound states after the quench of QW (include quenching parameters of the virtual and the real bulk system). Notice, during the quench process which is unitary, the  bulk topological invariant \citep{caio2015,d2015}, such as, Winding number \cite{McGinley2018}, Chern number \citep{d2015,caio2015,hu2016,wang2017} of the quantum state is unchanged. However, we can see that the number of bound states can be changed during the quench.

Unlike investigating the quenches in \citep{xu2020}, in which the system is prepared as the ground state of Hamiltonian $H_0$, here, the state before quench is the steady-state of a given $U(\theta_1,\theta_2)$ (with or without boundary state). Actually, in the semi-infinite system, we only focus on the left finite Fork states. The existence of the steady-state is justified by the probability $P(x<N_0)$ ($N_0$ is a given number), particularly $P_{\text{edge}}=p_0+p_1$. We focus on the time frame $R(\theta_{1}/2)S_{+}R(\theta_{2})S_{-}R(\theta_{1}/2)$ which preserves CS. When studying the quench of the real bulk system parameters, we control the parameter $\phi=0$ while changing the rotation angles $\theta_1,\theta_2$ in the experiment. The initial state is always prepared as $\left|0\right\rangle \otimes\left|\downarrow\right\rangle$.

Here, based on our experimental proposal, we carefully investigated the dynamics of the bound states at the boundary of the semi-infinite QW system. The investigation gives more information about the bulk-edge correspondence in non-equilibrium process.

\subsection{Formation of the bound state in the semi-infinite QW}

We first study the building up of the bound state when the initial state is prepared as $\left|0\right\rangle \otimes\left|\downarrow\right\rangle$. We study the dynamics of the system: with the fixed parameter $\theta_2=-\pi$ ($\phi=0$) in the virtual bulk system, and the selected rotation angle $(\theta_1,\theta_2)$ as $(\pi/2,0)$ (in the phase $(1,0)$), $(\pi/2,-2\pi/3)$ (in the phase $(0,0)$) and $(\pi/2,2\pi/3)$ (in the phase $(1,1)$ phase) for the real bulk system. To observe the establishment of the bound states, we monitor the evolution of the phonon state population of the walker in our experimental setup. In Fig. \ref{fig:bound state}a),  the edge population $P_{\text{edge}}$ is depicted, it is clear that when the topology of the real bulk system $(\theta_1=\pi/2, \theta_2=-2\pi/3)$ and the virtual bulk system ($\theta_1=\pi/2,\theta_2=-\pi$) are in the same phase $(0,0)$, the population will soon decay to zero (about 10 steps in our setup). When there exists a $0$ energy (or $\pi$ energy) bound state between the virtual and the real bulk system (as the parameter $(\pi/2,0)$ shown in Fig. \ref{fig:bound state}a)), the population will decay and stabilize to about 0.5 in current situation. However, when there exist two bound states ($0$-energy and $\pi$-energy) at the boundary, the edge population will also decay and stabilize to a nonzero value which is bigger than the situation where only one edge mode exists (see in Fig. \ref{fig:bound state}a)). 

We further fit the stable value of the population distribution of the phonon state in Fig. \ref{fig:phonon distribution}, and it can be well described by the exponential decay. The topology of the real bulk system ($\theta_1=\pi/2,\theta_2=0$, blue curve in Fig. \ref{fig:bound state}a)) in the phase $(1,0)$ are different from the virtual bulk system ($\theta_1=\pi/2,\theta_2=-\pi$). In this condition, we solved the eigenstate of evolution operator \cite{Kitagawa2012} and analytically find the localization length as $\lambda=1/\log (\sqrt{2}-1)$. The simulation results are given in Fig. \ref{fig:phonon distribution}a), b) for the evolution after 30 and 50 steps. The orange curve for the plot of the exponential decay with the localization length $\lambda$ mentioned above. We find that the fitting is pretty good. Particularly, $p_1/p_0$ and $p_2/p_1$ are exactly equal to $e^{-2/\lambda}$ during the evolution. We notice that the right moving bulk state contributes to the increase of mean and variation of the phonon distribution after each steps as in Fig. \ref{fig:phonon distribution}c). 

\begin{figure}
\includegraphics[width=8.5cm]{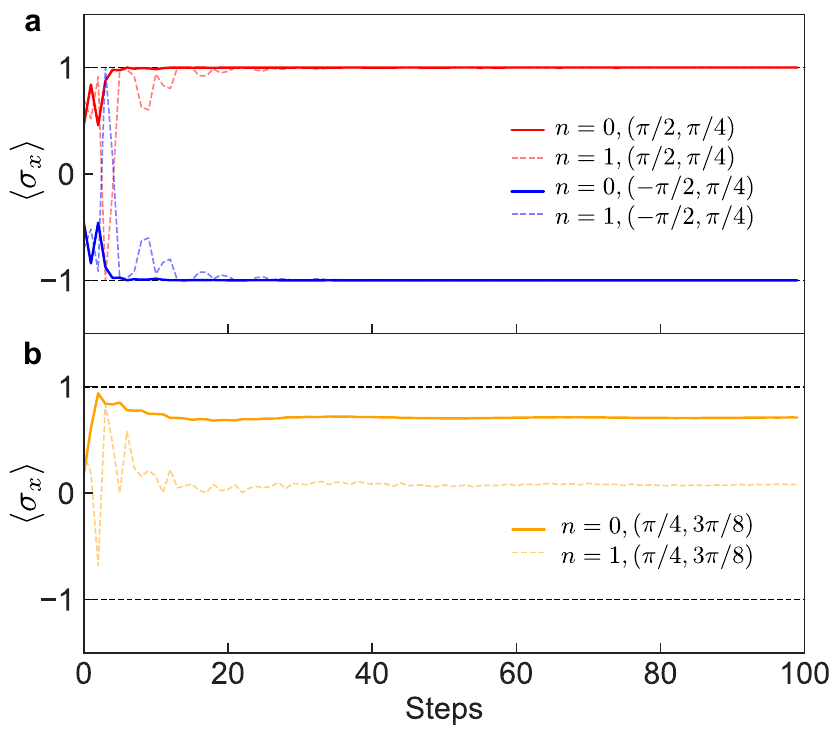}\caption{\label{fig:spin} Verify phases with the spin state of the quantum state $\left|\psi\right\rangle_{N}$. The particle is prepared as $\left|0\right\rangle \otimes\left|\downarrow\right\rangle$ and evolve after 100 steps. a. the spin state space evolution for the system with single bound state ($E=0,\pi$) and $n=0,1$ sites. Red (blue) lines correspond to a single 0 $(\pi)$-energy bound state. Average value of $\sigma_x$ stabilizes to 1 (-1). b. the spin state space evolution for the $(1,1)$ system as the superposition of 0 and $\pi$-energy bound state. Average value of $\sigma_x$ stabilizes to a value between 1 and -1 as expected, which are different for $n=0,1$ site.}
\end{figure}

Here, we would like to discuss how to experimentally verify different bound states appearing in the previous situations. Without loss of generality, we take $\phi=0$ in our simulation and suppose the bound state $\left|\psi_{b}\right\rangle$ is the eigenstate of evolution operator $U_{x\geq 0}$, i.e.
\begin{equation}
U_{x \geq 0}\left|\psi\right\rangle=e^{-i E}\left|\psi\right\rangle,
\end{equation}
with eigenergy $E$. Generally, the quantum state of the system after $N$ steps can be written as $\left|\psi\right\rangle_N=\sum_{n}(a_n(N)\left|\uparrow\right\rangle+b_n(N)\left|\downarrow\right\rangle)\otimes\left|n\right\rangle$ where all of $a_n(N), b_n(N)$ are real due to PHS. When the bound state is built (it is stable), that is, $\left|\psi\right\rangle_N =\left|\psi\right\rangle_{N+1}$ for $E=0$ while $\left|\psi\right\rangle_N = -\left|\psi\right\rangle_{N+1}$ for $E=\pi$, where  $\left|\psi\right\rangle_{N}$ is the wavefunction after $N$ steps ($N$ is large enough). We only focus on the evolution of the parameters $a_0$ and $b_0$ located on the boundary, i.e.

\begin{widetext}
\begin{equation}
	\begin{array}{l}
U(\theta_1,\theta_2)\left(\begin{array}{l}
a_{0}(N) \\
b_{0}(N)
\end{array}\right) \\
\rightarrow R \left(\frac{\theta_{1}}{2}\right)\left(\begin{array}{l}
\sin \left(\frac{\theta_{1}}{2}\right) a_{0}(N)+\cos \left(\frac{\theta_{1}}{2}\right) b_{0}(N) \\
\sin \left(\frac{\theta_{2}}{2}\right)\left(\cos \left(\frac{\theta_{1}}{2}\right) a_{0}(N)-\sin \left(\frac{\theta_{1}}{2}\right) b_{0}(N)\right)-\cos \left(\frac{\theta_{2}}{2}\right)\left(\sin \left(\frac{\theta_{1}}{2}\right) a_{1}(N)+\cos \left(\frac{\theta_{1}}{2}\right) b_{1}(N)\right)
\end{array}\right) \\
=\left(\begin{array}{l}
a_{0}(N+1) \\
b_{0}(N+1)
\end{array}\right).
\end{array}
\end{equation}
\end{widetext}
With the stable condition: $a_0(N)=a_0(N+1)$ and $b_0(N)=b_0(N+1)$, we can obtain:
\begin{equation}
    \begin{cases}
a_{0}=b_{0} & E=0,\\
a_{0}=-b_{0} & E=\pi. 
\end{cases}
\end{equation}
which is independent on the parameters $\theta_1$ and $\theta_2$ (however, the existence, the type of the bound states and the localization length are dependent on the parameters). As a result, when the bound state is the $0$-energy ($\pi$-energy) type, the spin state of the walker at the boundary is $\left|+\right\rangle=\frac{1}{\sqrt{2}}(\left|\uparrow
\right\rangle+\left|\downarrow\right\rangle)$ ($\left|-\right\rangle=\frac{1}{\sqrt{2}}(\left|\uparrow\right\rangle-\left|\downarrow\right\rangle)$) or $\langle \sigma_x \rangle=1$ ($\langle \sigma_x \rangle=-1$). Consequently, the $0$- and $\pi$-energy bound states are orthogonal and can be directly verified by measuring the operator $\sigma_x$. 

In Fig. \ref{fig:spin}, we simulate the evolution of the spin state (at $n=0,1$ site) in the different real bulk systems with the stable bound state. In Fig. \ref{fig:spin}a), the rotation angles of the real bulk system $(\pi/2,\pi/4)$ (in the phase $(1,0)$) and $(-\pi/2,\pi/4)$ (in the phase $(0,1)$) have different topology from the virtual bulk system ($(\pi/2,-\pi)$ and $(-\pi/2,-\pi)$). Red (blue) line for 0 ($\pi$)-energy bound state stabilize to $\left|+\right\rangle$ ($\left|-\right\rangle$). While for $(\pi/4,3\pi/8)$ (in the phase $(1,1)$), we find instead of stabilize to a fixed state, the spin state oscillates between two spin states ($a\left|+\right\rangle+b\left|-\right\rangle$ and $a\left|+\right\rangle-b\left|-\right\rangle$ for the two adjacent steps), which corresponds to a fixed $\langle \sigma_x \rangle$ between -1 and 1 as in Fig. \ref{fig:spin}b). 

In conclusion, we can easily verify the existence of the bound states by the edge population of the Fock states and further determined the type of the bound states by the average value of $\sigma_x$ in the spin state space: $\langle \sigma_x \rangle=1$ for $0$-energy bound state; $\langle \sigma_x \rangle=-1$ for $\pi$-energy bound state; $-1<\langle \sigma_x \rangle<1$ for the superposition of $0$-energy and $\pi$-energy bound state. The $0$- and $\pi$-energy bound state are product state $|\phi_s\rangle \otimes |\phi_p\rangle$ where $|\phi_s\rangle$ is the quantum state in the spin state space and $|\phi_p\rangle$ is the quantum state with exponential decay population distribution in the position space.   

\subsection{Dynamics of bound states in quenches of QW}

\begin{figure}
\includegraphics[width=8.5cm]{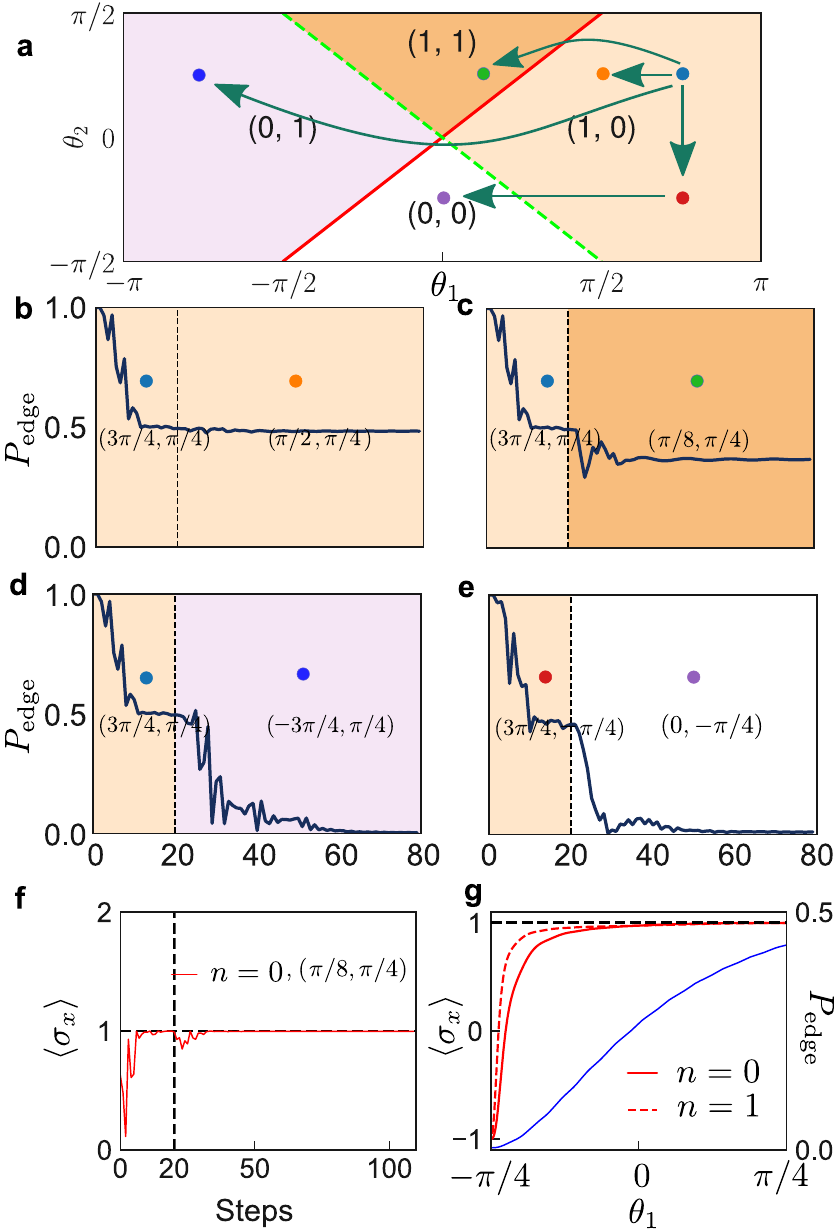}\caption{\label{fig:0quench} Quench dynamics starts from $(1,0)$ phase. The particle is prepared as $\left|0\right\rangle \otimes\left|\downarrow\right\rangle$. After first $N_0=20$ steps for building the bound state then the real bulk system suddenly quench to another phase. a. the parameters we study before and after the quench for the real bulk system. b. c. d. e. edge population
$P_{\text{edge}}$ verse evolution steps with the final real bulk system in the phases $(1,0)$, $(1,1)$, $(0,1)$, $(0,0)$ respectively. We notice that in all these cases, only the parameters after the quench in the phases $(1,1)$ and $(1,0)$ have none-zero bound state preserved. f. the spin state dynamics at $n=0$ site with parameters in c.: first a single $0$-energy bound state with $\left\langle\sigma_x\right\rangle=1$ is built at $N_0=20$ step, then oscillate and final still stabilize to $\left\langle\sigma_x\right\rangle=1$. g. expectation value $\left\langle\sigma_x\right\rangle$ and edge population $P_\text{edge}$ verse $\theta_1^f\in [-\pi/4,\pi/4]$ and fixed $\theta_2^f=\pi/4$, which ensure the system in the phase $(1,1)$ after the quench.}
\end{figure}

\begin{figure}
\includegraphics[width=8.5cm]{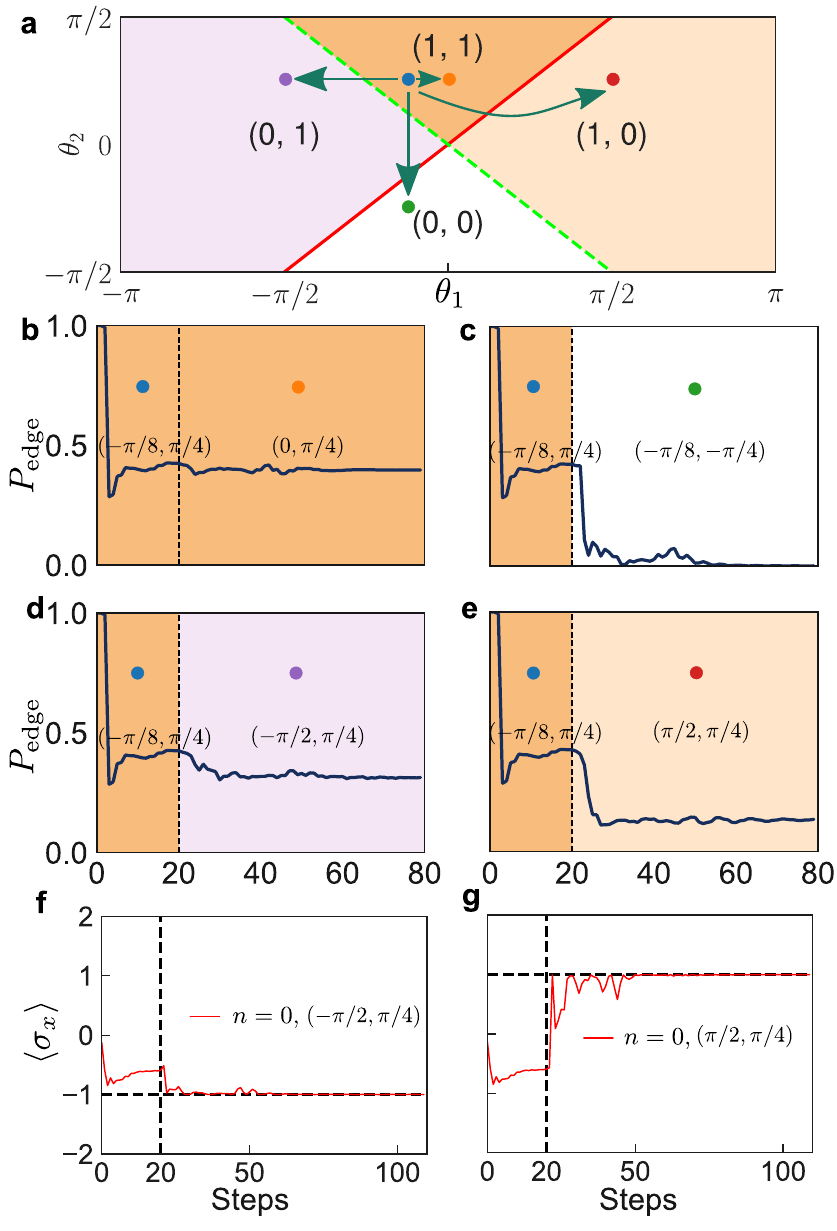}\caption{\label{fig:0piquench} Quench dynamics starts from $(1,1)$ phase. The particle is prepared as $\left|0\right\rangle \otimes\left|\downarrow\right\rangle$. After first $N_0=20$ steps for building the bound state then the real bulk system suddenly quench to another phase.  a. the parameters we study before and after the quench for the real bulk system.  b. c. d. e. edge population $P_{\text{edge}}$ verse evolution steps with the final real bulk system in the phases $(1,1)$, $(0,0)$, $(1,0)$, $(0,1)$ respectively. The bound can be preserved except the system in the phase $(0,0)$ after the quench. e(f). the spin state dynamics at $n=0$ site with the parameters after the quench in c(d). The spin state stabilize to $\left|+\right\rangle$ ($\left|-\right\rangle$) which shows a single 0 ($\pi$)-energy bound state.}
\end{figure}

\begin{figure}
\includegraphics[width=8.5cm]{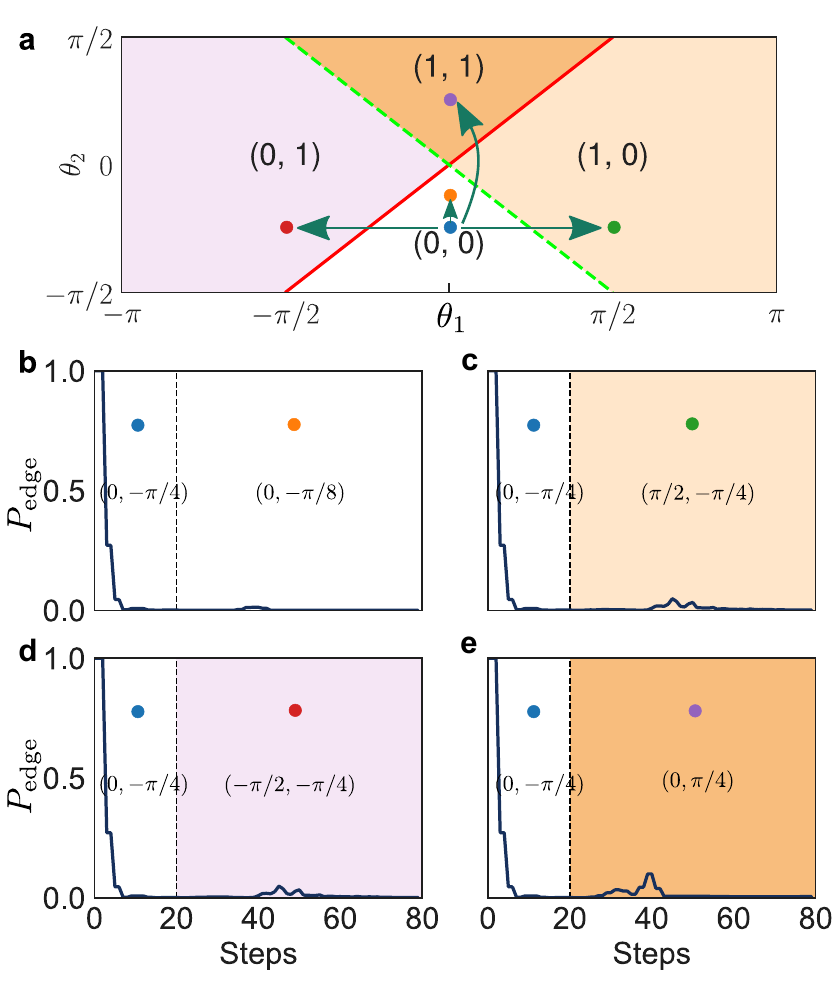}\caption{\label{fig:noquench} Quench dynamics starts from $(0,0)$ phase. The particle is prepared as $\left|0\right\rangle \otimes\left|\downarrow\right\rangle$. After the first $N_0=20$ steps, the real bulk system suddenly quench to another phase. a. the parameters we study before and after the quench for the real bulk system.  b. c. d. e. edge population $P_{\text{edge}}$ verse evolution steps with the final real bulk system in the phases $(1,1)$, $(0,0)$, $(1,0)$, $(0,1)$ respectively. We notice no bound state can be built when the system in the phase $(0,0)$ before the quench .}
\end{figure}

\begin{figure}
\includegraphics[width=8.5cm]{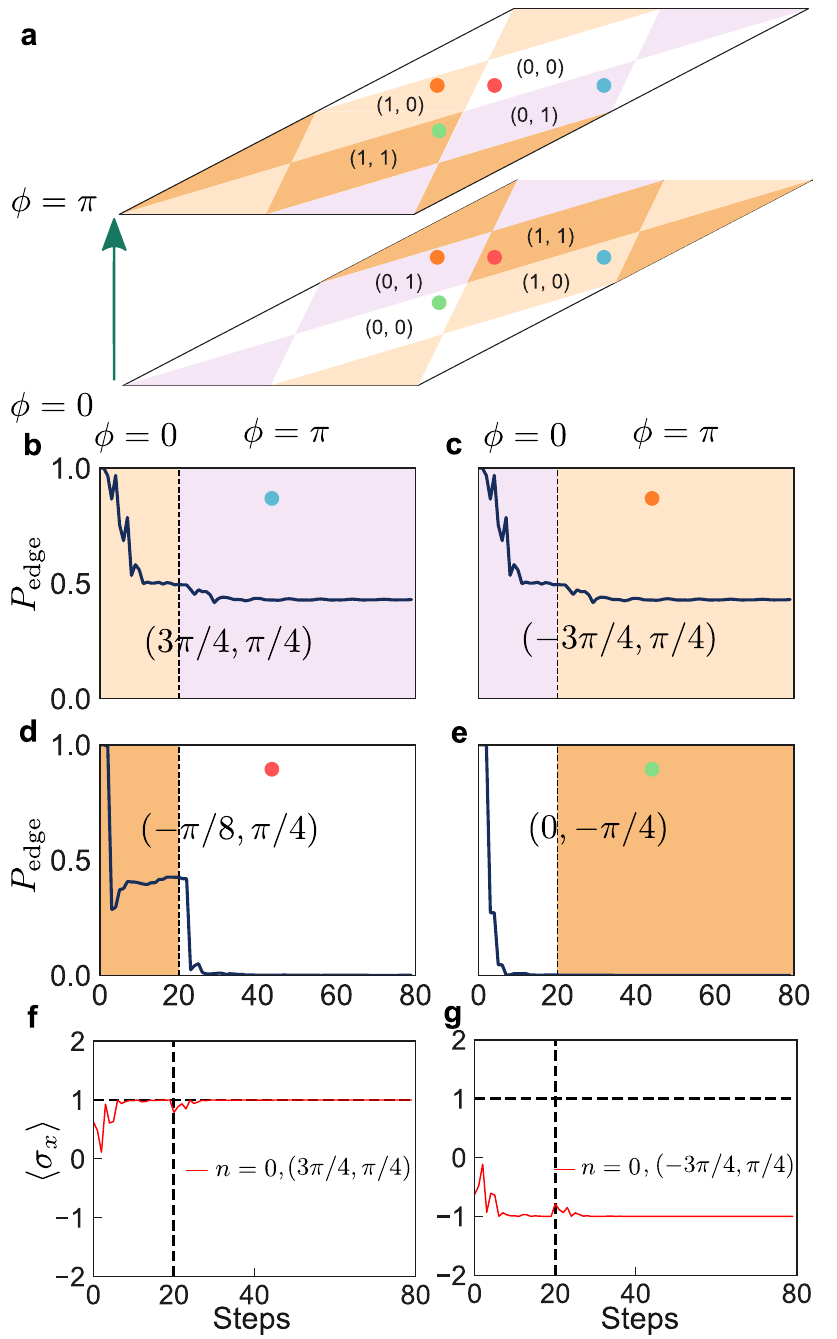}\caption{\label{fig:vquench} Quench dynamics of the the virtual bulk system. The particle is prepared as $\left|0\right\rangle\otimes\left|\downarrow\right\rangle$. a. diagram of quench the virtual bulk system. The system first evolve under the control parameter $\phi=0$ for $N_0=20$ steps, then suddenly quench $\phi$ to $\pi$ thus quench the virtual bulk system. b. c. d. e. edge population $P_{\text{edge}}$ verse evolution steps system before the quench in the phase $(1,0)$, $(0,1)$, $(1,1)$, $(0,0)$ and the system after the quench in the phase $(0,1)$, $(1,0)$, $(0,0)$, $(1,1)$ respectively. We notice in a. and b. bound states are preserved, and further study spin state dynamics under these two condition in e. and f.}
\end{figure}

\begin{figure}
\includegraphics[width=8.5cm]{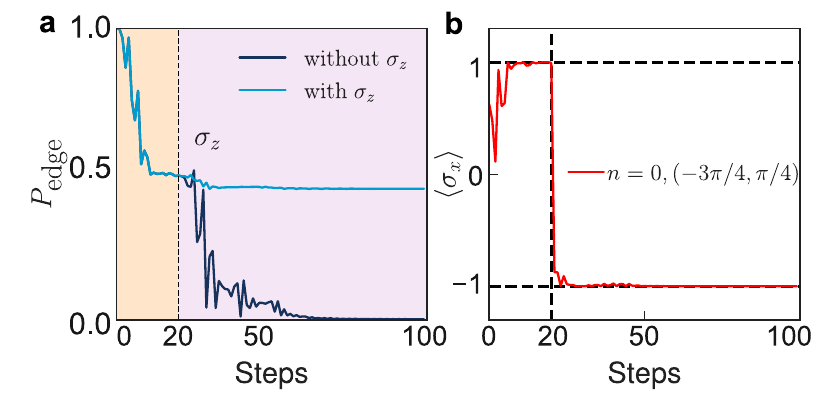}\caption{\label{fig:sigma}  Recovering of the bound state when quench between $(1,0)$ and $(0,1)$ with the same parameters in Fig. \ref{fig:0quench}d). a. edge population $P_{\text{edge}}$ verse evolution steps with and without $\sigma_z$ operation at the end of step $N_0$, which transfer $\left|+\right\rangle$ to $\left|-\right\rangle$, thus help recover the bound state. b. the spin state dynamics with the additional $\sigma_z$ operation. We can observe how a $\pi$-energy bound state is built starting from a $0$-energy bound state.}
\end{figure}

Now we turn to study the dynamics of the bound state in the quench processes. We study the sudden quench of rotation angles in the experiment and thus the real bulk system. The control parameter are set as $\phi=0$. Initially, the real bulk system is chosen as $(\theta_1^i,\theta_2^i)$, at the same time, the parameters of the virtual bulk system are $(\theta_1^i,-\pi)$. As mentioned before, if the real and the virtual bulk systems have different topology, the bound states will appear. After establishing the bound state by $N_0$-step evolution from the initial state $\left|0\right\rangle \otimes\left|\downarrow\right\rangle$, the rotation angles in the experiment changes at $N+1$ step, which means the real bulk system sudden quench to $(\theta_1^f,\theta_2^f)$ (the parameter of the virtual bulk system quenches to $(\theta_1^f,-\pi)$ simultaneously). Then the whole system evolved with these new parameters. Obviously, whether existence of the new bound states is strongly dependent on the parameters $\theta_1^f$ and $\theta_2^f$. Here, we only focus on the dynamics of the bound states by monitoring its edge population $P_{\text{edge}}$ still with left finite Fock states and the expectation value $\langle \sigma_x \rangle$ with the zero phonon state $\left|n=0\right\rangle $. 

Firstly, the initial rotation parameters of the real bulk system is $(3\pi/4,\pi/4)$ (in phase $(1,0)$) and the virtual bulk system is $(3\pi/4,-\pi)$ (in phase $(0,0)$) as shown in Fig. \ref{fig:0quench}a). Because the real and the virtual bulk system have different topology, after evolving $N_0=20$ steps, a $0$-energy bound state appears and becomes stable. Then, we change the rotation angles at 21 step thus sudden quench the real bulk system to different phases and maintain the virtual bulk system in the phase $(0,0)$. In Fig. \ref{fig:0quench}b), the real bulk system quenches to $(\pi/2,\pi/4)$ which is still in the phase $(1,0)$ (there is a $0$-energy bound state exist in the corresponding static system) and the virtual bulk system to $(\pi/2,-\pi)$. As a result, the bound state will be preserved without oscillation and decay. In Fig. \ref{fig:0quench}c), the real bulk system quenches to $(\pi/8,\pi/4)$ which is in phase $(1,1)$ and the virtual bulk system to $(\pi/8,-\pi)$. We can see in Fig. \ref{fig:0quench}b) that the edge population $P_{\text{edge}}$ will oscillate and decay first and stabilize to a new value in the end, which indicates the survival of the $0$-energy bound state. 

To make it clear, we further monitor the spin dynamics of zero phonon state  to determine the type of the bound state ($0$-energy or $\pi$-energy). As in Fig. \ref{fig:0quench}f), after the stabilization of the edge population, $\langle \sigma_x \rangle=1$ ($0$-energy bound state), i. e. the spin state of the bound state tends to be $\left|+\right\rangle$ indicates the bound state is $0$-energy. After the real bulk system quenches to the phase $(1,1)$, the expectation value, $\langle \sigma_x \rangle$ is still equal to 1 which indicates that the bound state is $0$-energy. Fig. \ref{fig:0quench}g) shows $\langle \sigma_x \rangle$ (red curve) and $P_{\text{edge}}$ (blue curve) verse parameters $(\theta_1,\pi/4)$ after quench. Here $\theta_1 \in [-\pi/4,\pi/4]$ to ensure the real bulk system after quench in the phase $(1,1)$.

In Fig. \ref{fig:0quench}d), the real bulk system quenches to $(-3\pi/4,\pi/4)$ which is in the phase $(0,1)$ and the virtual bulk system in the phase $(-3\pi/4,-\pi)$. A single $\pi$-energy bound state will exist with the parameters after the quench. While in Fig. \ref{fig:0quench}e), the real bulk system quenches to $(0,\pi/4)$ which is in the phase $(0,0)$ and the virtual bulk system in the phase $(0,-\pi)$.  In these two conditions mentioned above, we see that the edge population $P_{\text{edge}}$ will oscillate and finally decay to zero.

Secondly, we study the quench process starting from the establishment of the bound state with the real bulk system $(-\pi/8,\pi/4)$ (in the phase $(1,1)$) and the virtual bulk system $(-\pi/8,-\pi)$. Both the $0$-energy and $\pi$-energy bound state can exist before quench. The superposition of the $0$-energy and $\pi$-energy bound state can be verified by $\langle \sigma_x \rangle \neq 1$ in Fig. \ref{fig:0piquench}e) and f). Then, the real bulk system quech to the different phase while the virtual bulk system still in the phase $(0,0)$. In Fig. \ref{fig:0piquench}a), the parameter of the real bulk system quench to $(\pi/8,\pi/4)$ (also in the phase $(1,1)$) and the virtual bulk system in  $(\pi/8,-\pi)$. The bound state will be preserved after the quench dynamics and the population $P_{\text{edge}}$ almost the same. 

In Fig. \ref{fig:0piquench}b), the parameter of the real bulk system quenches to $(-\pi/8,-\pi/4)$ which is in the phase $(0,0)$ and the virtual bulk system in the phase $(-\pi/8,-\pi)$. Generally, there is no bound state in the quenched static system since the real and the virtual bulk system have the same topology. We can see that the bound state will quickly decay and disappear (the edge population soon decay to zero). In Fig. \ref{fig:0piquench}c) (d)), the real bulk system quenched to $(-\pi/2,\pi/4)$ (in phase $(0,1)$) ($(\pi/2,\pi/4)$ (in phase $(1,0)$)) and the virtual bulk system in the phase $(-\pi/2,-\pi)$ ($(\pi/2,-\pi)$). We can see the the survival of the bound state for the above conditions.

To see the detail of the survived bound states, we monitor their spin dynamics (see Fig. \ref{fig:0piquench}e) (f))). We find that the expection value $\langle \sigma_x \rangle$ will approach 1 (-1) in Fig. \ref{fig:0piquench}e) (f)) which indicates that only $0$-energy ($\pi$-energy) bound state survives and the other bound state decays in this situation.

Thirdly, we study the initial the real bulk system in phase $(0,0)$ with $(0,-\pi/4)$  and the virtual bulk system is also in phase $(0,0)$ with $(0,-\pi)$. There is no bound state in this system, and the edge population will approach to 0 after evolving the system $N_0=20$ steps. The real bulk system quenches to the different phase and the virtual bulk system keep in phase $(0,0)$: in Fig. \ref{fig:noquench}a), the real bulk system quenches to phase $(0,0)$ with parameter $(0,-\pi/8)$. In Fig. \ref{fig:noquench}b), the real bulk system quenches to phase $(1,0)$ with parameter $(\pi/2,-\pi/4)$; in Fig. \ref{fig:noquench}c), the real bulk system quench phase $(0,1)$ with parameter $(-\pi/2,-\pi/4)$; in Fig. \ref{fig:noquench}d), the real bulk system quench phase $(1,1)$ with parameter $(0,\pi/4)$ (in the phase $(1,1)$). For all the cases above, the bound state can be established.

Finally, we consider the quench of the virtual bulk system and keep the real bulk system parameters unchanged. As mentioned before, when the real bulk system is fixed by $(\theta_1,\theta_2)$, the virtual bulk system also can be controlled (tuned by the parameter $\phi$). As shown in Fig. \ref{fig:vquench}a), the parameter $\phi$ in the virtual bulk system before quench is set to $0$. The system supports the non-trivial bound state if the real bulk system $(\theta_1,\theta_2)$ has the different topology from the virtual bulk system. After the stabilization of the edge population (with enough evolution steps, $N_0=20$ in our simulation), the parameter $\phi$ of the virtual bulk system suddenly quenches to $\pi$. Then the system evolve with the parameter after the quench. We notice that with fixed $(\theta_1,\theta_2)$, and quench control parameter $\phi$ from $0$ to $\pi$, the original $(1,0)$ ($(0,1)$) phase turns to $(0,1)$ ($(1,0)$) phase, the original $(1,1)$ ($(0,0)$) phase turns to $(0,0)$ ($(1,1)$) phase.

In Fig. \ref{fig:vquench}b) ((c), the real bulk system is chosen in phase $(1,0)$ with parameter $(3\pi/4,\pi/4)$ (in the phase $(0,1)$ with parameter $(-3\pi/4,\pi/4)$), and the virtual bulk system is initially in phase $(0,0)$. There is a $0$-energy ($\pi$-energy) bound state located between these two bulk systems. After the quench, the real bulk system is in the phase $(0,1)$ ($(1,0)$), and the quenched virtual bulk system in $(3\pi/4,0)$ ($(-3\pi/4,0)$). The quenched real and virtual bulk systems still have different topology. The edge population $P_{\text{edge}}$ further evolve to the stable value.

Similarly, we monitor their spin dynamics in Fig. \ref{fig:vquench}e) (f)) to see the detail of the survived bound states in Fig. \ref{fig:vquench}a) (b)). We find that the the $\langle \sigma_x \rangle$ keep 1 (-1) in Fig. \ref{fig:0piquench}e) (f)) unchanged. That's because the single $0$-energy ($\pi$-energy) bound state under $\phi=0$ before quench becomes $\pi$-energy ($0$-energy) bound state under $\phi=\pi$ after quench, which can be survived in the system all the time. 

In Fig. \ref{fig:vquench}d) the real bulk system is in the phase $(1,1)$ with parameters $(-\pi/8,\pi/4)$ and the virtual bulk system is also initially in the phase $(0,0)$ with parameters $(-\pi/8,-\pi)$. 0 and $\pi$-energy bound state will be stable at the boundary. The virtual bulk system will quench to the original phase $(1,1)$ with parameter $(-\pi/8,0)$. The quenched virtual and the real bulk systems have the same topology and there is no bound state. Therefore, the bound state is quickly decaying and disappear. In Fig. \ref{fig:vquench}e) the real bulk system is in the phase $(0,0)$ with parameter $(0,\pi/4)$), and the virtual bulk system is also in the phase $(0,0)$ with parameters $(0,-\pi)$). They have the same topology and there is no bound state stable at the boundary. Similarly, the virtual bulk system will quench to the original phase $(1,1)$ and has the different topology from the real bulk system, however, the bound state can not be established in this situation.

With the previous simulation results, we can found that only both of the systems, before and after the quench, support the same type of the bound states, the bound state can exist after the quenching.  

In addition, we found that the $0$-energy and $\pi$-energy bound states have similar Fock states distribution, however, they have a different spin states when they are located at the boundary. Consequently, we can transfer the bound state with a very simple operator: implementing $\sigma_z$ on site $0$. As shown in Fig. \ref{fig:sigma}a) the blue curve with the $\sigma_z$ operation has a non-vanishing edge population $P_{\text{edge}}$ comparing with the purple curve. The spin dynamics in Fig. \ref{fig:sigma}b) further enhance our argument: first, the spin state of the bound state tend to be $\left|+\right\rangle$ indicates the $0$-energy bound state is built, then with $\sigma_z$ operation swap $\langle \sigma_x \rangle$ to -1 (thus $\left|-\right\rangle$). Then the bound state can further evolve and becomes stable in the phase $(0,1)$. 

\begin{figure}
\includegraphics[width=8.5cm]{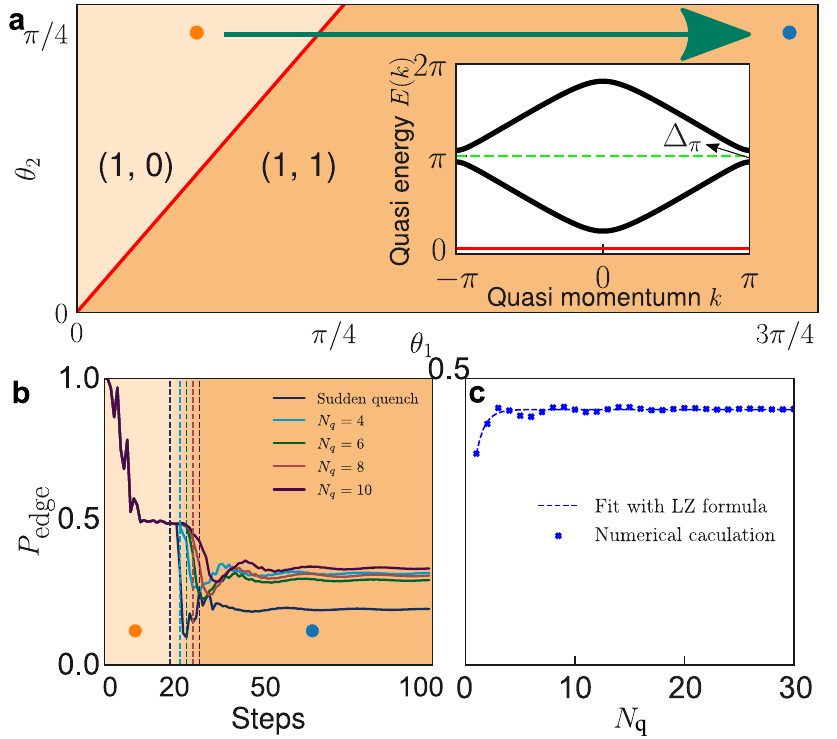}\caption{\label{fig:speed} Quench dynamics starts from $(1,0)$ phase with different quench steps $N_q$. The particle is prepared as $\left|0\right\rangle \otimes\left|\downarrow\right\rangle$.  a. the parameters we study before and after the quench for the real bulk system (same as in Fig. \ref{fig:0quench}c)). Inside: the energy dispersion spectrum in momentum space, red solid line for $E=0$ bound state and green dashed line for $E=\pi$ bound state. The quasi-energy gap $\Delta_{\pi}$ at $E=\pi$ set the quench steps scale jumping to the $\pi$-energy bound state. b. edge population $P_{\text{edge}}$ verse evolution steps with different $N_q$. c. stable state edge population $P_{\text{edge}}$ after the quench. Blue points for the numerical calculation while the blue dashed line fitting the result with the Landau-Zener formula, which has the form $e^{-\beta N_q}$ in our condition. $\beta$ is fitted to be 1.3.}
\end{figure}

\subsection{Edge population with different quench rates}

We further investigate the relation between the value of $P_{\text{edge}}$ and the quench rate. To clearly describe the effect of the quench rate, the initial parameters $\theta^i$ quenches to the final parameter $\theta^f$ by $N_q$ step and the time-dependent parameter $\theta^i$ has the form:

\begin{equation}
\theta(t)=\begin{cases}
\theta^{i} & t < N_0\\
\theta^{i} + \frac{\theta^{f}-\theta^{i}}{N_q}(t-N_0) & N_0\leq t \leq N_0+N_q\\
\theta^{f} & t > N_0+N_q
\end{cases},
\end{equation}
the first $N_0$ steps used to build the bound state (if any) between the real bulk system with parameter $(\theta_{1}^{i},\theta_2^{i})$ and the virtual bulk system with parameter $(\theta_{1}^{i},-\pi)$. Then, $N_q$ steps used to quench the real bulk system and the quench speed $v=\frac{\theta^{f}-\theta^{i}}{N_q}$, obviously, the larger $N_q$ is, the slower the quench dynamics happen. Finally, additional steps (after $N_0+N_q$ steps) are used to build the bound state (if any) between the real bulk system with parameter $(\theta_{1}^{f},\theta_2^{f})$ and the virtual bulk system with parameter $(\theta_1^{f},-\pi)$. In this case, we only investigate the relation between  $P_{\text{edge}}$ and the quench rate. All the proposals mentioned above can be realized precisely in the experiment by changing the rotation angles during each cycle.

In Fig. \ref{fig:speed}, we study the quench dynamics with different quench steps $N_q$. All of the quenches are, from the real bulk system in phase $(1,0)$ with parameter $(\pi/8,\pi/4)$ to phase $(1,1)$ with parameter $(3\pi/4,\pi/4)$, and the virtual bulk system are in the phase $(0,0)$. We thus have quench speed $v_q=\frac{5\pi}{8N_q}$. In the initial system, a single $0$-energy bound state can be stable after $N_0=20$ steps since the topology of the real and the virtual bulk system is different. In Fig. \ref{fig:speed}b), we study edge population verse evolution steps. Blue curves are for sudden quench ($N_q=1$), cyan curves are for $N_q=4$, green curves are for $N_q=6$, red curves for $N_q=8$ and purple curves are for $N_q=10$. We can see the population $P_{\text{edge}}$ is bigger when the quench is slower. When the quench is slow enough, $P_{\text{edge}}$ is almost the same as the $P_{\text{edge}}$ before the quench which indicates there is no transport happen. However, when the quench is quick, the population decays. This phenomena can be well understand by the Landau Process \citep{zener1932}. The bound state is isolated from the transport mode with a gap as shown in Fig. \ref{fig:vquench}a) inside, if the quench is fast, the isolated bound state has some probability to jump to the transport mode; however, if the quench is slow enough, the process is almost adiabatic, and the bound mode can not jump to the transport modes. To further understand this process, we fit the population $P_{\text{edge}}$ V.S. $N_q$, which can be well described by the Landau process as $e^{-\alpha\Delta_{\pi}^2/v_q}$  (thus $e^{-\beta N}$) in Fig. \ref{fig:speed}a) inside. Intuitively, when the $v_q<\Delta_{\pi}$, there will be a limiting possibility of jumping to the $\pi$-energy bound state. $v_q=\frac{5\pi}{8N_q}=\Delta_{\pi}$ gives $N_q\approx10$ coin with Fig. \ref{fig:speed}c): with $N_q>10$, $P_{\text{edge}}$ is almost the same as the $P_{\text{edge}}$ before the quench.

\section{CONCLUSION\label{sec:6}}

In this paper, we have proposed a proposal to realize QW in the Fock states with carefully designed laser sequence in a trapped ion. In this proposal, the properties and the dynamics of the bound states can be experimentally observed with the natural boundary. Particularly, the quench dynamics of the bound states with energy $0$ or $\pi$ can be monitored by the population of the phonon states and the expectation value of the operator $\left\langle\sigma_x\right\rangle$ of the selected internal level of the ion. Different quench dynamics have been comprehensively discussed, with the development of the manipulation of the phonons \cite{zhang2018} in a trapped ion, all the required techniques are available and it can be realized currently.

\bibliographystyle{apsrev4-1}
\phantomsection\addcontentsline{toc}{section}{\refname}\nocite{*}
\bibliography{qw}

\end{document}